\documentclass[aps,prl,twocolumn,superscriptaddress,amsfonts]{revtex4-1}

\usepackage{graphicx}
\usepackage{amssymb}
\usepackage{amsmath}

\usepackage[colorlinks=true,citecolor=blue,linkcolor=magenta]{hyperref}
\hypersetup{
pdfauthor = {P. Maletinsky},
colorlinks = true, linkcolor = blue, urlcolor=blue, bookmarksnumbered =  true}

\newcommand{\ket}[1]{\left|#1\right>}
\newcommand{\CRI}{CrI$_3$}
\newcommand{\vdw}{vdW}

\usepackage{graphicx}

\begin{document}

\title{Probing magnetism in 2D materials at the nanoscale with single spin microscopy} 

\author{L.~Thiel}
\affiliation{Department of Physics, University of Basel, Klingelbergstrasse 82, Basel CH-4056, Switzerland}
\author{Z.~Wang}
\affiliation{Department of Quantum Matter Physics, University of Geneva, 24 Quai Ernest Ansermet, CH-1211 Geneva, Switzerland}
\affiliation{Group of Applied Physics, University of Geneva, 24 Quai Ernest Ansermet, CH-1211 Geneva, Switzerland}
\author{M.~A.~Tschudin}
\affiliation{Department of Physics, University of Basel, Klingelbergstrasse 82, Basel CH-4056, Switzerland}
\author{D.~Rohner}
\affiliation{Department of Physics, University of Basel, Klingelbergstrasse 82, Basel CH-4056, Switzerland}
\author{I.~Guti\'errez-Lezama}
\affiliation{Department of Quantum Matter Physics, University of Geneva, 24 Quai Ernest Ansermet, CH-1211 Geneva, Switzerland}
\affiliation{Group of Applied Physics, University of Geneva, 24 Quai Ernest Ansermet, CH-1211 Geneva, Switzerland}
\author{N.~Ubrig}
\affiliation{Department of Quantum Matter Physics, University of Geneva, 24 Quai Ernest Ansermet, CH-1211 Geneva, Switzerland}
\affiliation{Group of Applied Physics, University of Geneva, 24 Quai Ernest Ansermet, CH-1211 Geneva, Switzerland}
\author{M.~Gibertini}
\affiliation{Department of Quantum Matter Physics, University of Geneva, 24 Quai Ernest Ansermet, CH-1211 Geneva, Switzerland}
\affiliation{National Centre for Computational Design and Discovery of Novel Materials (MARVEL), \'Ecole Polytechnique F\'ed\'erale de Lausanne, CH-1015 Lausanne, Switzerland}
\author{E.~Giannini}
\affiliation{Department of Quantum Matter Physics, University of Geneva, 24 Quai Ernest Ansermet, CH-1211 Geneva, Switzerland}
\author{A.~F.~Morpurgo}
\affiliation{Department of Quantum Matter Physics, University of Geneva, 24 Quai Ernest Ansermet, CH-1211 Geneva, Switzerland}
\affiliation{Group of Applied Physics, University of Geneva, 24 Quai Ernest Ansermet, CH-1211 Geneva, Switzerland}
\author{P.~Maletinsky}
\email{patrick.maletinsky@unibas.ch}
\affiliation{Department of Physics, University of Basel, Klingelbergstrasse 82, Basel CH-4056, Switzerland}

\date{\today}


\maketitle

\textbf{\color{black}
The recent discovery of ferromagnetism in 2D van der Waals (\vdw{}) crystals has generated widespread interest, owing to their potential for fundamental and applied research. 
Advancing the understanding and applications of \vdw{} magnets requires methods to quantitatively probe their magnetic properties on the nanoscale. 
Here, we report the study of atomically thin crystals of the \vdw{} magnet \CRI{} down to individual monolayers using scanning single-spin magnetometry, and demonstrate quantitative, nanoscale imaging of magnetisation, localised defects and magnetic domains. 
We determine the magnetisation of \CRI{} monolayers to be $\approx16~\mu_B/$nm$^2$ and find comparable values in samples with odd numbers of layers, whereas the magnetisation vanishes when the number of layers is even.  
We also establish that this inscrutable even-odd effect is intimately connected to the material structure, and that structural modifications can induce switching between ferro- and anti-ferromagnetic interlayer ordering. 
Besides revealing new aspects of magnetism in atomically thin \CRI{} crystals, these results demonstrate the power of single-spin scanning magnetometry for the study of magnetism in 2D \vdw{} magnets. 
}

\begin{figure}[t!]
\includegraphics[width=8.6cm]{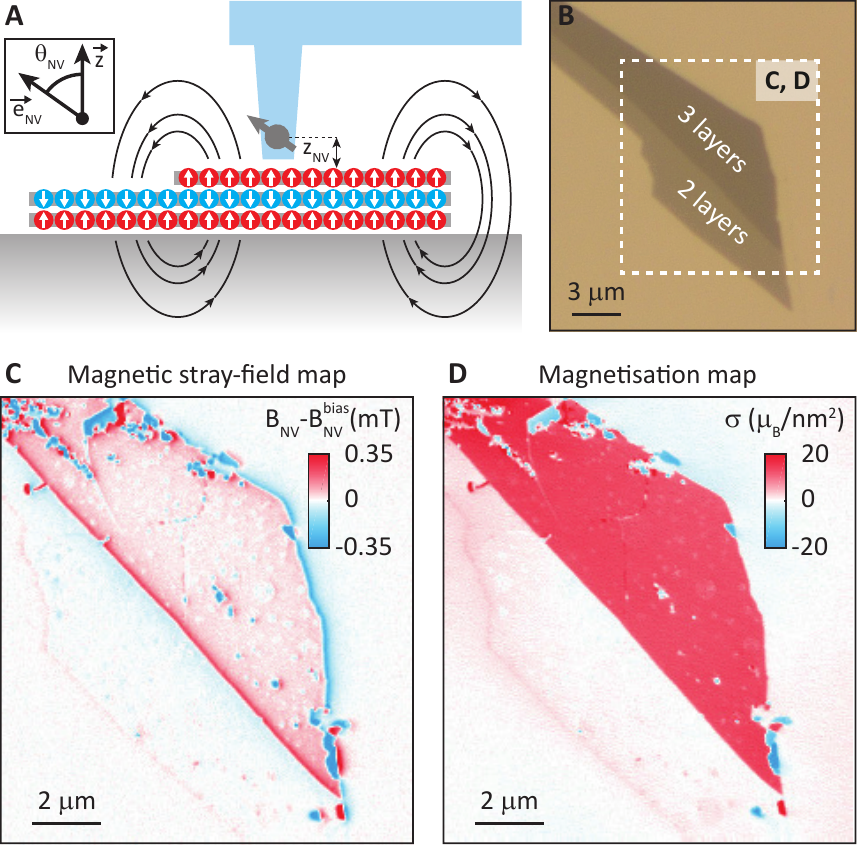}
\caption{\label{LblFig1} {\bf Nanoscale imaging of magnetism in two-dimensional van der Waals magnets.} 
{\bf A}~Schematic of the scanning single spin magnetometry technique employed in this work. 
A single Nitrogen-Vacancy (NV) electronic spin is scanned across few layer flakes of encapsulated CrI$_3$ (encapsulation not shown). 
Stray magnetic fields from the sample are sensed by optically detected Zeeman shifts of the NV spin states, and imaged with nanoscale resolution (set by the sensor-sample separation $z_{\rm NV}$) by lateral scanning of the NV. 
The method detects magnetic fields along the NV spin quantisation axis $\vec{e}_{\rm NV}$, at an angle $\theta_{\rm NV}~\sim54^\circ$ from the sample normal.
{\bf B}~Optical micrograph of the \CRI{} bi-and tri-layer flake of sample D1.  
{\bf C}~Magnetic field map of $B_{\rm NV}$ across sample D1 recorded in a bias field $B_{\rm NV}^{\rm bias}=172.5~$mT and at a typical green laser power $P_{\rm laser}\approx40~\mu$W. Strong (close to zero) stray fields emerge from the edges of the trilayer (bilayer) flake, respectively. 
{\bf D}~Map of \CRI{}'s magnetisation distribution in sample D1, determined by unique reverse-propagation of the magnetic field map in {\bf C} (see text and\,\cite{SOM}). 
}\end{figure} 

\begin{figure*}[t!]
\includegraphics[]{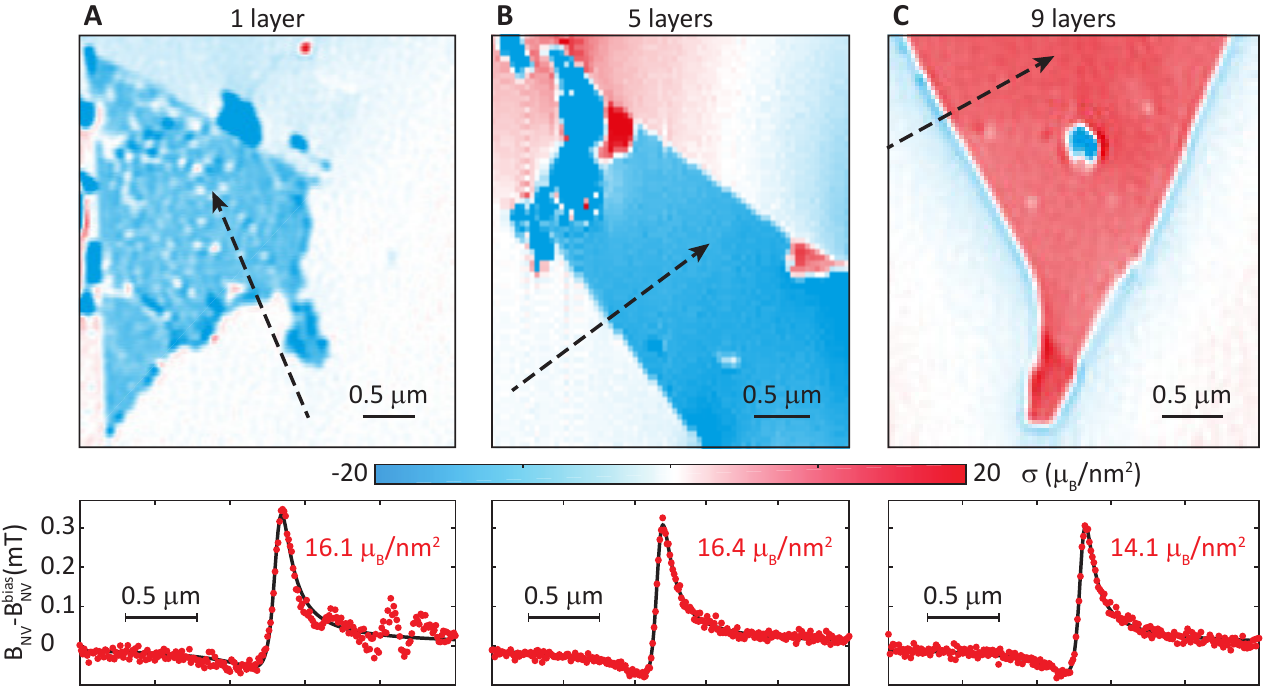}
\caption{\label{LblFig2} {\bf Magnetisation maps of few-layer \CRI{} flakes}. 
Data were acquired after zero field cooling and spontaneous magnetic ordering, and under experimental conditions described in Fig.\,\ref{LblFig1}. 
Magnetisation maps were obtained for a monolayer ({\bf A}), a five ({\bf B}) and a nine ({\bf C}) layer thick sample (see text and\,\cite{SOM}). 
The colorbar applies to all panels. 
Lower row: Independently acquired data of magnetic field $B_{\rm NV}$ measured across the borders of each flake, along the lines indicated in the maps. 
These data allow for an independent determination of magnetisation $\sigma_z$ and sensor-sample separation $z_{\rm NV}$ using analytic fits (black)\,\cite{Hingant2015}.
}\end{figure*}

Magnetism in individual monolayers of \vdw{} crystals has recently been observed in a range of materials, 
including semiconducting\,\cite{Huang2017a,Gong2017a} and metallic\,\cite{Bonilla2018,Fei2018a,Deng2018a} compounds.
The discovery of such two dimensional magnetic order is per se non-trivial\,\cite{Mermin1966} and has triggered significant attention owing to emerging exotic phenomena including Kitaev spin liquids\,\cite{Banerjee2016a,Banerjee2017a}, or novel magneto-electric effects\,\cite{Wang2018b,Jiang2018b,Jiang2018a,Huang2018a}. Remarkable efforts have led to the use of two-dimensional magnets as functional elements in spintronics, such as spin-filters\,\cite{Song2018a,Klein2018}, spin-transistors\,\cite{Jiang2018c}, tunnelling magnetoresistance devices\,\cite{Wang2018a,Kim2018a} or magneto-electric switches\,\cite{Jiang2018b,Jiang2018a,Huang2018a}. Further advances hinge on methods for the quantitative study of the magnetic response of these atomically thin crystals at the nanoscale, but despite their central importance, the required experimental methods are still lacking. Indeed, transport experiments\,\cite{Song2018a,Jiang2018c,Klein2018,Wang2018a,Jiang2018b,Jiang2018a} probe magnetic properties only indirectly. The only existing, spatially resolved studies rely on optical techniques, such as fluorescence\,\cite{Zhong2017,Seyler2018a} or the magneto-optical Kerr effect (MOKE)\,\cite{Huang2017a,Gong2017a,Fei2018a}, and are therefore limited to the micron-scale. Even more critically, these techniques do not provide quantitative information about the magnetisation (MOKE, e.g., may yield non-zero signals even for antiferromagnets\,\cite{Sivadas2016}) and are susceptible to interference effects that can obscure magnetic signals in thin samples\,\cite{Huang2017a}.

In this work, we overcome these limitations and present a powerful approach for quantitative addressing of nanoscale magnetic properties of \vdw{} magnets, which we here illustrate on the prominent case of \CRI{}. Specifically, we employ a scanning Nitrogen-Vacancy (NV) centre spin in diamond as a sensitive, atomic-scale magnetometer\,\cite{Rondin2014} to quantitatively determine key magnetic properties of \CRI{} and to directly image magnetic domains with spatial resolutions of few tens of nanometres. For magnetometry, we exploit the Zeeman effect, which leads to a shift of the NV spin's energy levels as a function of magnetic field. In the regime relevant for this work, the NV spin shows a linear Zeeman response for magnetic fields $B_{\rm NV}$ along its spin-quantisation axis $\vec{e}_{\rm NV}$ (Fig.\,\ref{LblFig1}{\bf A}), while it is largely insensitive to fields orthogonal to $\vec{e}_{\rm NV}$\,\cite{Rondin2014}. The NV spin therefore offers a direct and quantitative measurement of the vectorial component $B_{\rm NV}$ of stray magnetic fields emerging from a sample. 

The Zeeman shifts of the involved NV spin-levels can be conveniently read out by optically detected electron spin resonance (ODMR) using $532~$nm laser excitation, microwave spin driving and NV fluorescence detection for spin readout\,\cite{Gruber1997,SOM}. 
For nanoscale imaging, we employ a single NV spin held in the tip of an atomic force microscope and approach the NV to within a distance $z_{\rm NV}\sim60~$nm to the sample, which then results in a magnetic imaging resolution on the order of $z_{\rm NV}$\,\cite{Rondin2014}. The scanning probe containing the NV\,\cite{Maletinsky2012, Appel2016} is integrated into a confocal optical microscope for optical spin readout and the whole apparatus immersed in a liquid-$^4$He cryostat. Superconducting magnets are used to enable vectorial magnetic field control up to $0.5~$T. The measurement temperature of $\sim7~$K was determined by a resistive thermometer placed close to the sample.

We studied \CRI{} samples of various thicknesses, which were encapsulated in either h-BN or graphene to assure the stability of \CRI{} under oxygen atmosphere (for details, see\,\cite{SOM}). The samples (Fig.\,\ref{LblFig1}{\bf B}) were fabricated by mechanical exfoliation of \CRI{} and subsequently encapsulated using an established pick-and-place technique described elsewhere\,\cite{Wang2018a}. Samples of various \CRI{} thicknesses in the range of few ($<10$) layers were produced to study the effect of thickness on magnetic ordering. For each \CRI{} sample, the number of atomic layers was determined by a combination of AFM and optical microscopy\,\cite{SOM}. To prepare and study the magnetic state of \CRI{}, the samples were mounted in the NV magnetometer and cooled in zero magnetic field to the final measurement temperature. 

Figure\,\ref{LblFig1}{\bf C} shows a typical magnetic field map we acquired on an area containing bilayer and trilayer \CRI{} (sample D1).
The presented data were acquired in a bias field $B_{\rm NV}^{\rm bias}=172.5~$mT, where magnetometer performance was optimal, but equivalent images and results were found at lower fields as well. We obtained such maps from NV ODMR spectra acquired at each pixel (acquisition time $\approx2~$s/pixel), from which we determined $B_{\rm NV}$ through a fit\,\cite{SOM}. We confirmed experimentally that for typical experimental parameters we employed, our approach induced no significant back-action onto the sample, e.g., through heating by laser illumination or microwave irradiation\,\cite{SOM}. The resulting data show stray magnetic fields emerging predominantly from the edges of the tri-layer flake, as expected for a largely uniform magnetisation\,\cite{Hingant2015}, and thereby provide clear evidence for the magnetisation of few-layer \CRI{} we seek to study. 

To reveal further details of the underlying magnetisation pattern in \CRI{}, we use well-established reverse propagation protocols\,\cite{Roth1989} to map the magnetic field image of $B_{\rm NV}$ to its source (see\,\cite{SOM} for details). For a two-dimensional, out-of-plane magnetisation, as in the present case, such reverse propagation yields a unique determination of the underlying magnetisation pattern (Fig.\,\ref{LblFig1}{\bf D}), provided that the distance $z_{\rm NV}$ between sensor and sample is known. In our refined reverse propagation protocol, we determine $z_{\rm NV}$ through an iterative procedure described elsewhere\,\cite{Appel2018,SOM}. The reverse propagation thereby additionally yields the spatial resolution of our images, which is directly given by $z_{\rm NV}$ (for the data in Fig.\,\ref{LblFig1}, $z_{\rm NV}=62~$nm). In the following, we will focus our discussion on magnetisation maps obtained through such reverse propagation, while the raw magnetic field images are presented in\,\cite{SOM}.

The magnetisation pattern in Fig.\,\ref{LblFig1}{\bf D} clearly shows a largely homogenous magnetisation for this trilayer \CRI{} flake, which is typical for most samples we investigated. In addition, sparsely scattered, localised defects, mostly with vanishing magnetisation, were visible across the flake and few irregularities occured at the flake edges, which are likely caused by curling and rippling induced on the edges during sample preparation. On the flake, we found an average magnetisation $\sigma_z\approx 13.0~\pm2.4~\mu_B/$nm$^2$ (with $\mu_B$ the Bohr magneton), consistent with a single layer of fully polarised Cr$^{+3}$ spins, for which $\sigma_z^{\rm mono}=14.7~\mu_B/$nm$^2$ would be expected\,\cite{McGuire2015a}. The data thus supports the notion of antiferromagnetic interlayer exchange coupling in few-layer \CRI{}\,\cite{Huang2017a}, which results in a net magnetisation $\sigma_z^{\rm mono}$ for the magnetically ordered trilayer sample.

Sample D1 additionally contained a region of bilayer \CRI{} for which we observed zero bulk moment, again consistent with antiferromagnetic interlayer coupling. However, the bilayer also showed a weak magnetisation located within less than our spatial resolution of its edge (Fig.\,\ref{LblFig1}{\bf D}). The origin of this magnetisation is currently unknown, but could be related to magneto-electric effects\,\cite{Wang2018b, Jiang2018a}, a narrow region of monolayer \CRI{} protruding from the bilayer, or to spin canting\,\cite{Garcia-Sanchez2014} close to the edge of the flake.

\begin{figure}[t!]
\includegraphics[width=8.6cm]{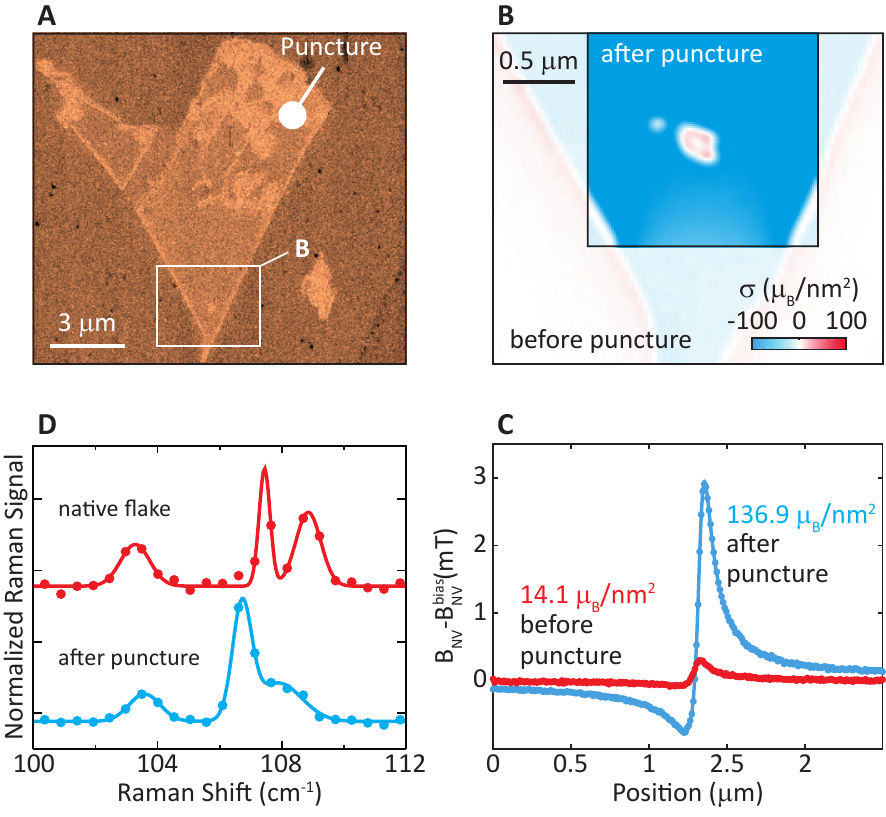}
\caption{\label{LblFig3}  {\bf Interplay of structural and magnetic order in few-layer \CRI{}}. 
{\bf A}~Overview image of sample D2 (see\,\cite{SOM}), indicating the location of the puncture, and the representative region sampled in the following. 
{\bf B}~Enhanced magnetisation observed in the 9-layer flake of sample D2, after puncturing the \CRI{} encapsulation. Note that the location of the puncture is few $\mu$m away from the imaged region (see {\bf A}). The faint background shows the (inverted) data from Fig.\,\ref{LblFig2}{\bf C} for comparison. The colorbar applies to all data.
{\bf C}~Line-cut of $B_{\rm NV}$ across a representative edge of the punctured flake (blue), indicating a magnetisation $\sigma_z=136.9\pm0.4~\mu_B/$nm$^2$, close to the expected value for a fully polarised nine-layer sample. For reference, the red data shows the corresponding data from before the puncture.
{\bf D}~Raman spectra from the punctured flake and an unpunctured reference, demonstrating the concurrence of a structural transition together with the magnetic transition. 
}
\end{figure}

We applied our measurement procedure to a variety of samples, including a monolayer and the 5- and 9-layer flakes  shown in Fig.\,\ref{LblFig2}{\bf A}, {\bf B} and {\bf C}, respectively. Strikingly, all these flakes exhibit near-uniform magnetisation at a magnitude comparable to $\sigma_z^{\rm mono}$.  We additionally determined $\sigma_z$ in an independent way by measuring $B_{\rm NV}$ along lines crossing the edges of each flake (Fig.\,\ref{LblFig2}, lower panels). Assuming a purely out-of-plane magnetisation, analytical fits\,\cite{Hingant2015} to these data allow for the quantitative determination of both $\sigma_z$ and $z_{\rm NV}$. \textcolor{black}{
Potential rotations of the magnetisation away from $z$ in the vicinity of the edge due to, e.g. the Dzyaloshinskii-Moriya interaction, would only lead to negligible deviations from our findings\,\cite{Tetienne2014b}. For the monolayer, 5-, and 9-layer flakes, we then find $\sigma_z=16.1\pm 0.6~\mu_B/$nm$^2$, $16.4\pm 0.2~\mu_B/$nm$^2$ and $14.1\pm 0.2~\mu_B/$nm$^2$, respectively, where uncertainties denote statistical errors of the fit ($z_{\rm NV}\approx60~$nm in all cases, see\,\cite{SOM}). The general agreement of these fits with the values of $\sigma_z$ found in Fig\,\ref{LblFig2}{\bf A}, {\bf B} and {\bf C} further confirms the validity of the reverse propagation method we employed.}

Our observations thus far corroborate previous results on few-layer \CRI{}\,\cite{Huang2018a,Seyler2018a,Song2018a},  which all found \CRI{} flakes with odd (even) numbers of layers to exhibit non-zero (close to zero) magnetisation as a result of antiferromagnetic interlayer exchange coupling. 
These observations, however, are in conflict with the established fact that \CRI{} is a bulk ferromagnet\,\cite{Dillon1965}. We shed light on this dichotomy in a subsequent experiment on sample D2, where our diamond scanning probe induced an  unintentional local puncture through the encapsulation layer of the \CRI{} flake (Fig.\,\ref{LblFig3}{\bf A} and\,\cite{SOM}). After this, the whole sample, up to several microns away from the puncture, exhibited a significantly enhanced magnetisation, as evidenced by a representative magnetisation map (Fig.\,\ref{LblFig3}{\bf B}) and  $B_{\rm NV}$ linescan (Fig.\,\ref{LblFig3}{\bf C}) across the flake. The data show a $\approx9.7$-fold increase of magnetisation from initially $14.1\pm0.2~\mu_B/$nm$^2$ to $136.9\pm 0.4~\mu_B/$nm$^2$. For the 9-layer flake under study, this enhancement suggests a transition from antiferromagnetic to ferromagnetic interlayer coupling induced by the puncture. 

To investigate the occurrence of a structural transition in our punctured sample we have compared in Fig.\,\ref{LblFig3}{\bf D} its low-temperature Raman spectrum with the one of a pristine flake, in a spectral region where characteristic Raman modes for \CRI{} exist\,\cite{Djurdjic2018a}. Although the data does not allow for an unambiguous determination of the crystalline structure of our samples, the markedly different spectra clearly point to a change in structure occurring simultaneously with the change in magnetic order discussed above. This observation is consistent with recent results of density functional calculations\,\cite{Jiang2018,Soriano2018,Sivadas2018,Jang2018a,Wang2018a}, predicting an interplay between stacking order and interlayer exchange coupling in \CRI{}. The nature of the structural transition induced by the puncture and the crystalline structure prior to the puncture of the flake need to be elucidated and will be the subject to further research.

\begin{figure}[t!]
\includegraphics[width=8.6cm]{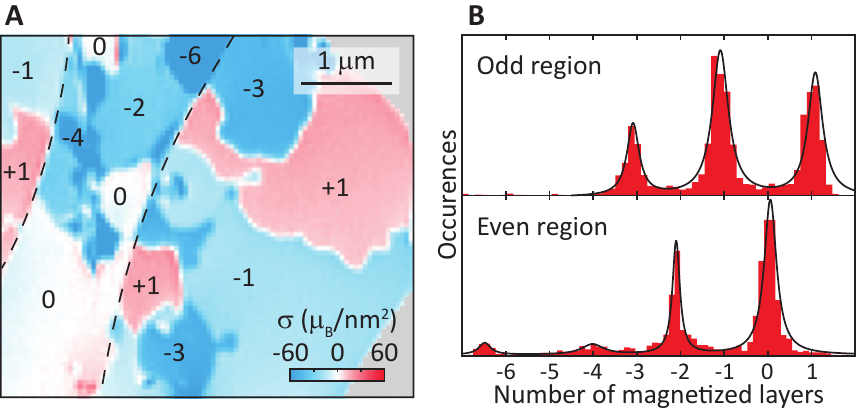}
\caption{\label{LblFig4} {\bf Magnetic domains in \CRI{}}. 
{\bf A}~Spontaneously occurring magnetic domains observed in the nine-layer sample D2. 
The magnetisation $\sigma_z$ was found to be discretised in integer multiples of the monolayer magnetisation $\sigma_z^{\rm mono}$. Numbers indicate $\sigma_z$ in multiples of $\sigma_z^{\rm mono}$, where positive (negative) values denote magnetisations along the $+z$ ($-z$) direction. A clear separation of the flake into regions of even- and odd-multiples of $\sigma_z^{\rm mono}$ is observed. Even-numbered regions are assigned to a missing or added monolayer (see text). Areas where no flake was present were false-color coded in grey. 
{\bf B}~Histograms of magnetisation pixel-values obtained in the odd- and even-numbered regions of the data in {\bf A}.
}\end{figure}

The connection between crystal structure and magnetism also offers an explanation for the occurrence of magnetic domains in some of our \CRI{} samples. Figure\,\ref{LblFig4}{\bf A} shows a representative image of such domains on sample D2 (9-layer flake). Strikingly, the measured domain magnetisations only assume values close to integer multiples of $\sigma_z^{\rm mono}$, i.e. $\sigma_z=n\sigma_z^{\rm mono}$, with $n\in\mathbb{Z}$. This observation can be explained by spatial variations (in all three dimensions) of exchange couplings, which may alter the ordering of pairs of \CRI{} layers from antiferromagnetic to ferromagnetic between adjacent domains, e.g. due to a local change in crystal structure or stacking order. While this domain formation mechanism would preserve the parity of $n$, we observe well-separated regions on the sample where $n$ is either even or odd (see outlines in Fig.\,\ref{LblFig4}{\bf A} and histogram in Fig.\,\ref{LblFig4}{\bf B}). The removal or addition of a monolayer of \CRI{} between these areas can explain this observation and could have occurred during material exfoliation or sample preparation.

We have employed scanning NV magnetometry to observe a direct connection between structural and nanoscale magnetic ordering in \CRI{}, and thereby address the key question of why few-layer \CRI{} shows antiferromagnetic interlayer exchange coupling despite the  bulk being ferromagnetic. 
Beyond \CRI{}, our work establishes scanning NV magnetometry as a unique tool to address nanoscale magnetism in \vdw{} crystals, down to the limit of a single atomic layer. 
Such direct, quantitative imaging and sensing is vital to further our fundamental understanding of these materials and their development towards applications in future spintronics devices\,\cite{Bonilla2018,Deng2018a}. 
Our approach is general and can even be applied under ambient conditions\,\cite{Rondin2014,Bonilla2018} or to materials where optical methods could not be applied thus far\,\cite{Ghazaryan2018a}. Finally, the ability to perform NV magnetometry on \vdw{} magnets offers perspectives for high-frequency sensing\,\cite{Du2017} of their magnonic excitations\,\cite{Ghazaryan2018a,Klein2018}, and thereby develop \vdw{} magnets towards novel, atomic-scale platforms in magnonics applications.

\section{Acknowledgements}
We thank A.~H\"ogele, M.~Munsch, and J.-V. Kim for fruitful discussions and valuable feedback on the manuscript and A. Ferreira for technical help. 
We gratefully acknowledge financial support from the SNI; NCCR QSIT; SNF grants 143697, 155845, 169016 and 178891 the EU Graphene Flagship. N.U. and M.G. gratefully acknowledge support through an Ambizione fellowship of the Swiss National Science Foundation. 

\section{Author contributions}
All authors contributed to all aspects of this work.

\section{Additional information}
The authors declare no competing financial interest. 


\bibliography{20190201_Cri3}

\clearpage
\pagebreak
\newcommand{\beginsupplement}{%
	\setcounter{table}{0}
	\renewcommand{\thetable}{S\arabic{table}}%
	\setcounter{figure}{0}
	\renewcommand{\thefigure}{S\arabic{figure}}%
}

\beginsupplement
\onecolumngrid

\begin{center}
	\large
	\textbf{Supplementary information for\\ ``Probing magnetism in 2D materials at the nanoscale with single spin microscopy''}
\end{center}

\normalsize

\section{Principle of isomagnetic field imaging}
\label{isofield}

\begin{figure}[h!]
	\includegraphics[width = 12cm]{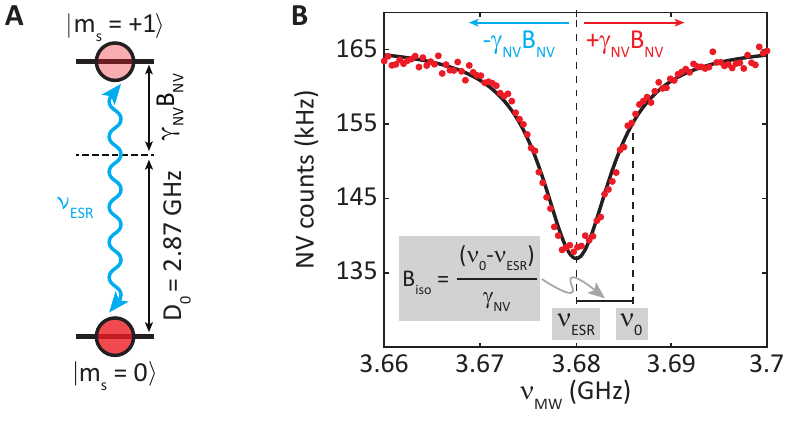}
	\caption{{\bf Principle of isomagnetic field imaging} {\bf A} Level scheme of the NV center with sublevels $\ket{m_s = 0}$ and $\ket{m_s = +1}$ split in energy by the zero field splitting $D_0=2.87~$GHz and an applied magnetic field $\gamma_{\rm NV}B_{\rm NV}$. While $\ket{m_s = 0}$ yields a high fluoresence rate, NV fluoresence is on average $20\%$ weaker upon optical excitation from $\ket{m_s = +1}$. {\bf B} For optically detected electron spin resonance, the NV center is initialized in $\ket{m_s = 0}$, while the frequency of a microwave driving field $\nu_{\rm MW}$ is swept over the NV transition at frequency $\nu_{\rm ESR}$. When $\nu_{\rm MW} = \nu_{\rm ESR}$, population from $\ket{m_s = 0}$ is transfered to $\ket{m_s = +1}$ leading to a dip in florescence. In an isomagnetic field scan the microwave frequency is fixed at $\nu_{0}$ and NV fluoresence will be minimal whenever it experiences a magnetic field of amplitude $B_{\rm iso} = (\nu_0-\nu_{\rm ESR})/\gamma_{\rm NV}$.}
	\label{Fig_ISOfield}
\end{figure}

The NV center orbital ground state forms an electronic S=1 spin triplet with magnetic sublevels $\ket{m_s = 0,\pm1}$ and a zero field splitting between  $\ket{m_s = 0}$ and  $\ket{m_s = \pm1}$ of $D_0=2.87~$GHz\cite{Rondin2014}. Figure\,\ref{Fig_ISOfield}A shows the two level system spanned by the $\ket{m_s = 0}$ and $\ket{m_s = +1}$ sublevels, which is used for magnetometry in this work. Upon application of an external magnetic field B$_{NV}$ along the NV-axis, the  $\ket{m_s = +1}$ sublevel experiences a Zeeman shift of
\begin{equation}
\Delta\nu = \gamma_{\rm NV}B_{\rm NV},
\end{equation} 
where $\gamma_{\rm NV}=28~$MHz/mT is the gyromagnetic ratio. As magnetic fields perpendicular to the NV axis have to compete with the zero field splitting $D_0$, they only yield a second order effect and, can therefore be ignored in this work.

Optical initizalization and read-out of the NV center\cite{Gruber1997} then allows for optical detected magnetic resonance (ODMR) where the NV center is initialized in its bright $\ket{m_s = 0}$ state and a mircowave driving field at frequency $\nu_{\rm MW}$ populates the less bright $\ket{m_s = +1}$ state in the resonance case $\nu_{\rm MW} = \nu_{\rm ESR}$, leading to a dip in fluoresence (Fig.\,\ref{Fig_ISOfield}B). The B$_{\rm NV}$ maps shown in the main text have been obtained by recording such ODMR spectra at each pixel of the scan at typical pixel dwell-times of $2~$s.

An alternative, faster method to acquire an overview of the magnetic signal originating from a sample is isomagnetic field imaging in which contours of fixed values B$_{\rm NV} = $B$_{\rm iso}$ are measured. To that end, the microwave frequency is fixed at $\nu_{0} = \nu_{\rm ESR}+\gamma_{NV}B_{\rm iso}$ and the sample is scanned below the NV center while its fluoresence is constantly interrogated. Whenever the sample stray field at the position of the NV corresponds to B$_{\rm iso}$, a decrease in fluoresence is observed, leading to magnetic contrast in the iso field image. This procedure was used to obtain the data in Fig.\,3A of the main text with B$_{\rm iso}=0$.

\newpage
\section{Encapsulation of micromechanical cleaved CrI$_3$ crystals}
\label{Nanofab}

The few-layer CrI$_3$ flakes studied here were micromechanically cleaved from chemical vapour deposition (CVD) grown CrI$_3$ single crystals and then transferred onto $90~$nm SiO$_2$/Si wafers inside a N$_2$ filled glovebox ($<1~$ppm of H$_2$O and $<20~$ppm of O$_2$). They were then encapsulated using either thin graphite ($3-10~$nm) or hBN ($5-20~$nm) flakes inside the same glove box to avoid degradation in air. Encapsulation was performed via a standard all-dry pick-up and release technique using a polymer stack\cite{Wang2013}.  

\begin{figure}[h!]
	\includegraphics[width = 10cm]{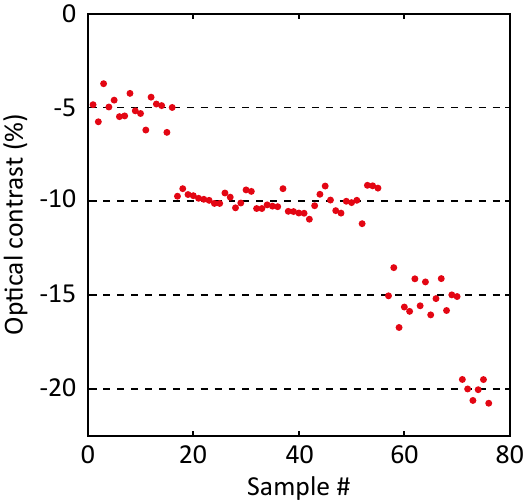}
	\caption{{\bf Relative optical contrast of atomically thin CrI3 crystals.} The contrast is calculated as $C = \frac{I_{\rm flake}-I_{\rm substrate}}{I_{\rm flake}+I_{\rm substrate}}$ where the $I_{\rm flake}$ ($I_{\rm substrate}$) is the reflected red-channel intensity from the flake (substrate). Contrast of around 5\%, 10\%, 15\% and 20\% correspond to mono-, bi-, tri- and tetra-layers CrI$_3$, respectively.}
	\label{Fig_OpticalContrast}
\end{figure}

\begin{figure}[h!]
	\includegraphics[width = 16.0cm]{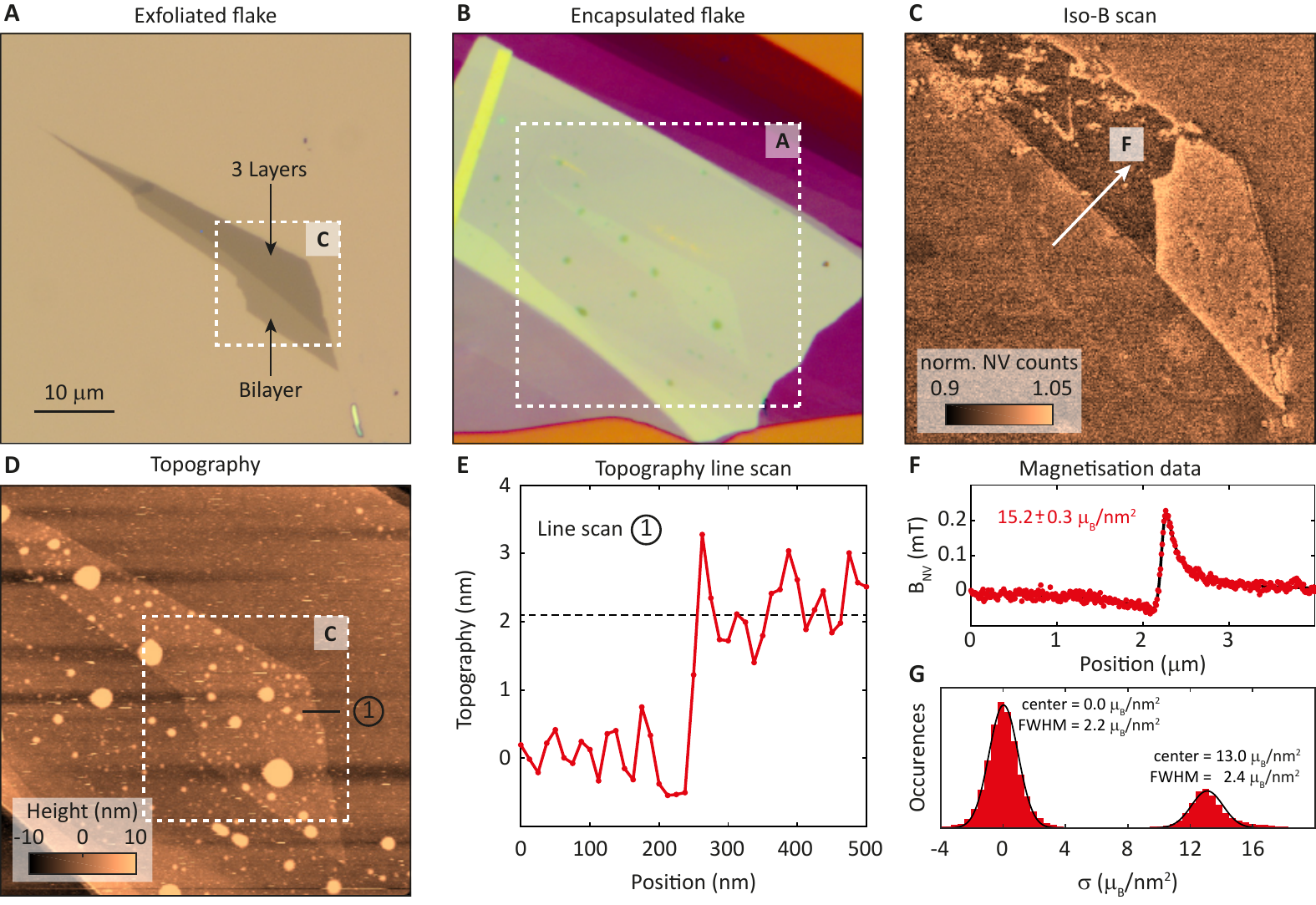}
	\caption{{\bf Characterization of sample D1 consisting of a bi- and trilayer of CrI$_3$.} {\bf A} Optical image of the exfoliated flake. {\bf B} Optical image after encapsulation with graphene. {\bf C} Iso-field image ($B_{\rm iso}=0.12~$mT) of the section illustrated with the white dashed box in {\bf A} and {\bf D}. Integration time is $0.2~$s per pixel. The three layer flake features a domain wall which was introduced by saturating the magnetisation in a negative magnetic field of $B_{\rm NV}=-495~$mT and subsequently applying a positive field of B$_{\rm NV}=170~$mT. {\bf D} Ex-situ AFM data of the encapsulated sample. {\bf E} Line scan of the topography over the three layer edge as indicated in {\bf D}. The theoretical height value for a three layer flake ($2.1~$nm) is displayed as a dashed line. {\bf F} Magnetic field of linecut over edge of trilayer flake with its extracted magnetisation is shown. The location of the linecut is indicated with a white arrow in {\bf C}. {\bf G} Histogram of the magnetisation in units of $\mu_B$/nm$^2$ of {\bf Fig.\,1D} in the main text. The histogram is fitted with a Gaussian function and the mean and full width at half max (FWHM) is displayed in the graph.}
	\label{Fig_Sample_2_3_layer}
\end{figure}

\begin{figure}[h!]
	\includegraphics[width = 16.0cm]{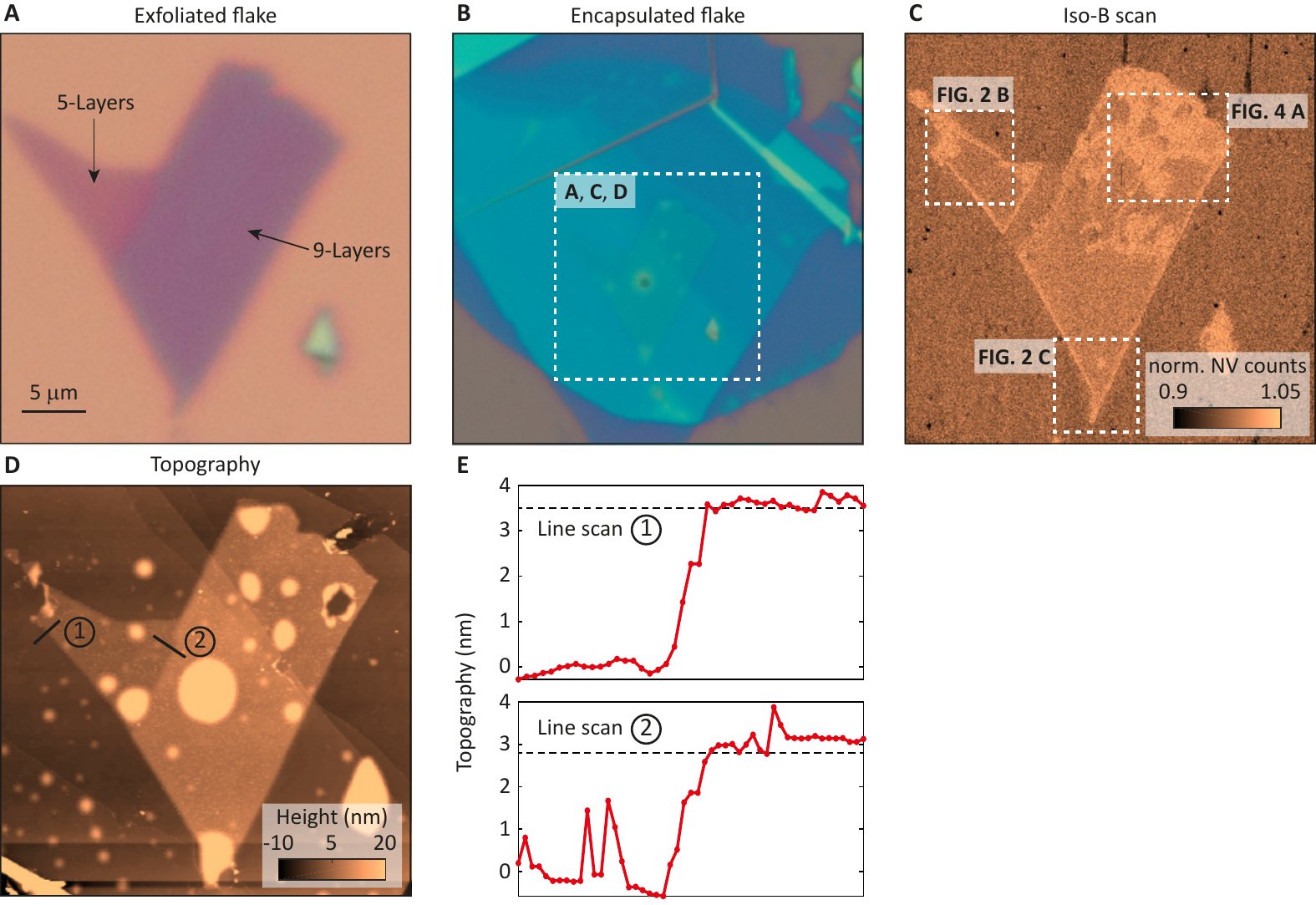}
	\caption{{\bf Characterization of sample D2 consisting of five and nine layer of CrI$_3$.} {\bf A} Optical image of the exfoliated flake. {\bf B} Optical image after encapsulation with hBN. {\bf C} Iso-field image ($B_{\rm iso}=0$) of the section illustrated with the white dashed box in {\bf B} before the puncture of the hole. Integration time was $0.18~$s per pixel. The scan sections of {\bf Fig.\,2B}, {\bf Fig.\,2C} and {\bf Fig.\,4A} in the main text are indicated. {\bf D} Ex-situ AFM data of the encapsulated sample after the puncture of the hole. {\bf E} The top panel shows the line scan of the topography over the five layer edge as indicated in {\bf D}. The theoretical height value for a five layer flake ($3.5~$nm) is displayed as a dashed line. The bottom panel shows the line scan of the topography over the five to nine layers edge as indicated in {\bf D}. The theoretical height value for four layers ($2.8~$nm) is displayed as a dashed line.}
	\label{Fig_Sample_5_9_layer}
\end{figure}

\begin{figure}[h!]
	\includegraphics[width = 16.0cm]{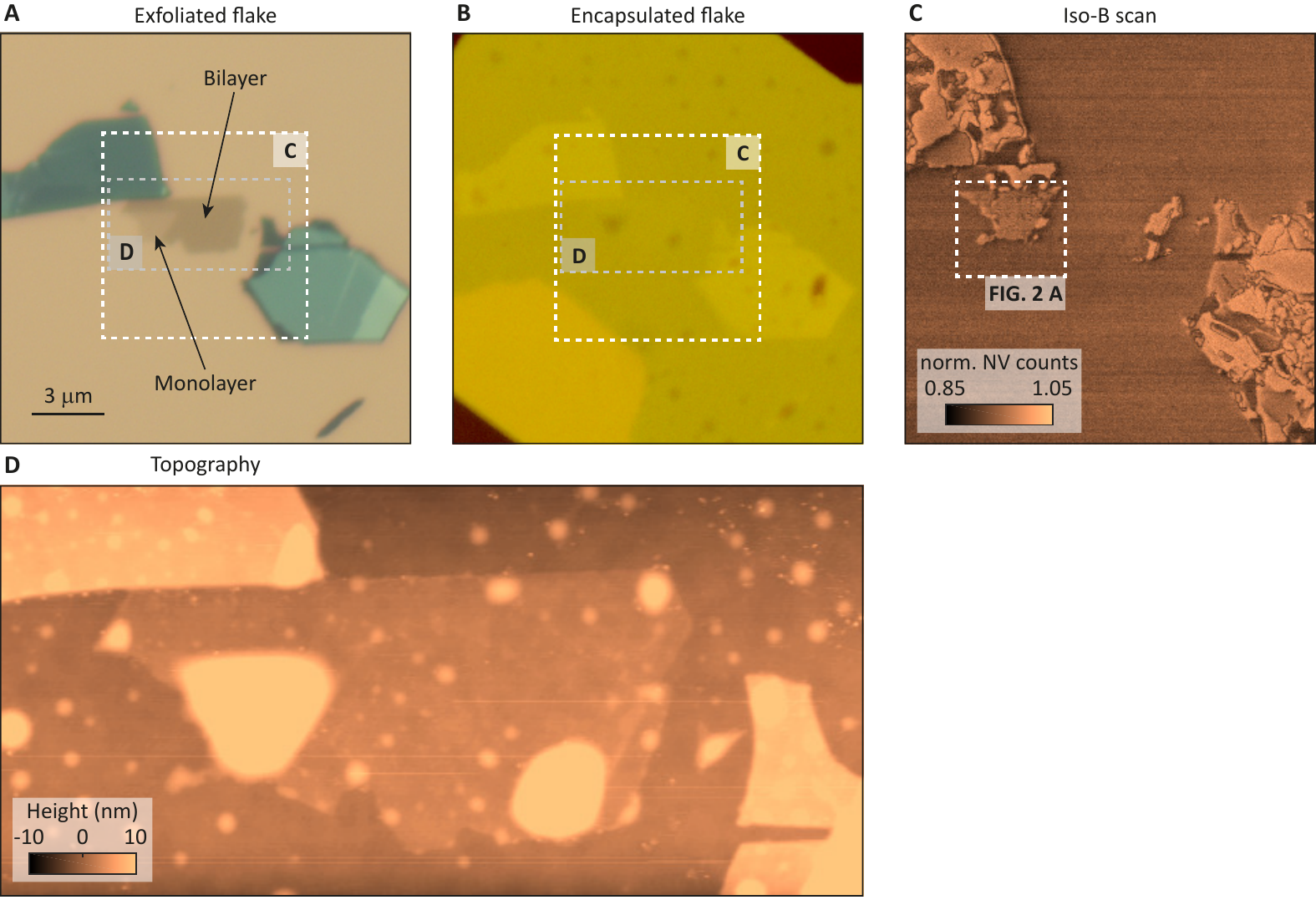}
	\caption{{\bf Characterization of sample D3 consisting of a mono- and bilayer of CrI$_3$.} {\bf A} Optical image of the exfoliated flake. {\bf B} Optical image after encapsulation with graphene. The same section of the sample is shown as in {\bf A}. {\bf C} Iso-field image ($B_{\rm iso}=0.27~$mT) of the section illustrated with the white dashed box in {\bf A} and {\bf B}. Integration time was $0.8~$s per pixel. Magnetic signal from the thick flakes as well as the monolayer sample is clearly visible while no magnetic signal emerges from the bilayer. The scan section of {\bf Fig.\,2A} in the main text is indicated. {\bf D} Ex-situ AFM data of the encapsulated sample of the section illustrated with the grey dashed box in {\bf A} and {\bf B}.}
	\label{Fig_SampleMonolayer}
\end{figure}

Figure\,\ref{Fig_Sample_2_3_layer}A, Fig.\,\ref{Fig_Sample_5_9_layer}A and Fig.\,\ref{Fig_SampleMonolayer}A show optical micrographs of the mono-, bi-, tri-, five- and nine-layers CrI$_3$ crystals whose magnetometry data is discussed in the main text, before encapsulation. Their thickness was determined using the relative optical contrast (red channel) between the CrI$_3$ crystals and the $90~$nm SiO$_2$ layer\cite{Huang2017a}, which was calibrated via the analysis of 80 atomically thin crystals (1 to 4 layers), as shown in Fig.\,\ref{Fig_OpticalContrast}. The thickness of the CrI$_3$ was confirmed by atomic force microscope measurements after the encapsulation.

\newpage
\section{Raman spectra of few-layer CrI$_3$ flakes}
\label{Raman}

The Raman scattering experiment was performed using a Horiba scientific (LabRAM HR Evolution) confocal microscope in backscattering geometry. After laser excitation the dispersed light was sent to a Czerni-Turner spectrometer equipped with a $1800~$groves/mm grating, which resolves the optical spectra with a precision of $0.3~$cm$^{-1}$. The light was detected with help of a thermopower cooled CCD-array. The excitation wavelength of the laser was $532~$nm. The samples were mounted in the cryostat (cryovac KONTI cryostat) with optical access. The Raman peaks were fitted using Lorentzian lineshapes.

\newpage
\section{Sample D1}
\label{SamplesD1}

\newpage
\section{Sample D2}
\label{SamplesD2}

\newpage
\section{Sample D3}
\label{SamplesD3}

\newpage
\section{Magnetic field data}
\label{BNV}

\begin{figure}[h!]
	\includegraphics[width = 12cm]{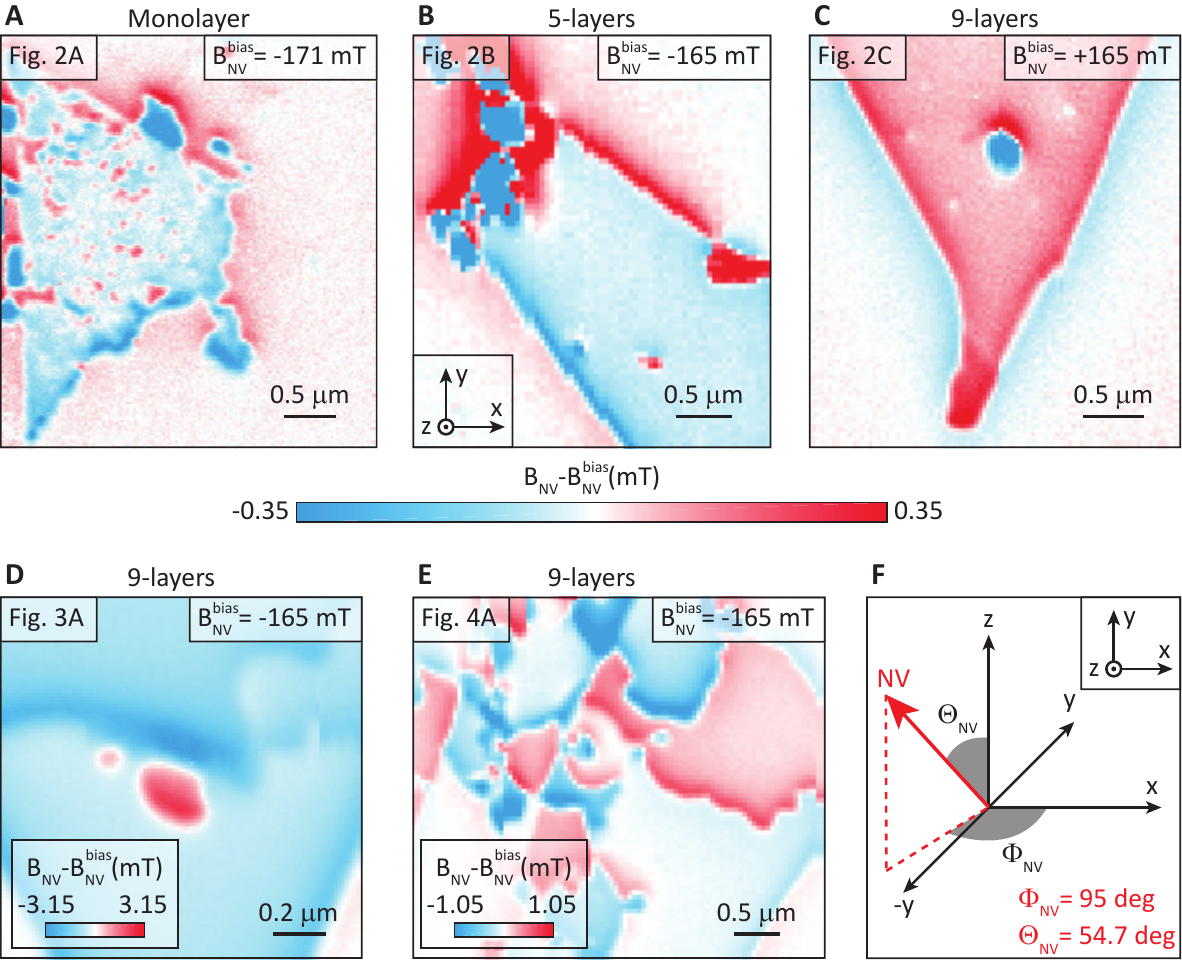}
	\caption{{\bf Data set of measured $B_{NV}$ maps.} {\bf A}-{\bf E} are maps of the stray magnetic field projected onto the NV-axis. The magnetisation maps shown in the main text are based on the magnetic field maps presented here. The corresponding figures in the main text and background magnetic fields $B_{\rm NV}^{\rm bias}$ are indicated in the figures themselves. The bias magnetic field perpendicular to the flake can be obtained from $B_z= B_{\rm NV}^{\rm bias}\cdot\cos(\Theta_{\rm NV}$) with $\Theta_{\rm NV}= 54.7^\circ$. {\bf F} Illustration of the NV axis on which the stray magnetic field is projected for {\bf A}-{\bf E}. The inset shows the coordination system for the field maps {\bf A}-{\bf E}. For all magnetisation maps shown in the main paper, $\Phi_{\rm NV}= 95^\circ$ and $\Theta_{\rm NV}= 54.7^\circ$ was used in the reverse propagation, since the same scanning NV spin was used for all experiments.}
	
	\label{Fig_MagneticFieldData}
\end{figure}

\newpage
\section{Reverse propagation of magnetisation}
\label{ReversePropagation}

We deployed a reverse propagation method to retrieve the perpedicular magnetisation M$_{\rm s}$(x,y,z=0) of the sample from our measured magnetic field map B$_{\rm NV}$(x,y,z=z$_{\rm 0}$)\cite{Hug2004}. The method is performed in Fourier space with Fourier-space coordinates ${\bf k} = (k_{\rm x}, k_{\rm y})$ in the $(x, y)$-plane.
The half-space above the sample contains no time-dependent electric fields or currents and therefore
\begin{equation}
\label{eq_rotH}
\nabla\times {\bf H} = 0. 
\end{equation}
Hence, we can define a magnetic potential $\phi_{\rm m}({\bf k},z)$ such that 
\begin{equation}
\label{eq_gradPhi}
{\bf H}=-\nabla~\phi_{\rm m},
\end{equation}
where $\nabla = (ik_{\rm x},ik_{\rm y},\partial/\partial z)$ is the gradient in Fourier space.

In order to fulfill Eq.(\ref{eq_rotH}), the Laplace equation $\Delta\phi_{\rm m} = 0$ needs to hold outside the sample, which leads to $\phi_{\rm m}({\bf k},z) = \phi_{\rm m}({\bf k},0)e^{-{\bf k}z}$ and $\nabla = (ik_{\rm x},ik_{\rm y},-k)$. For a purely out-of-plane magnetisation $H_{\rm z} = M_{\rm s}/2$ on the sample surface, and, using Eq.(\ref{eq_gradPhi}), we then obtain $\phi_{\rm m}({\bf k},z) = M_{\rm s}({\bf k},z)/(2k)$ and therefore
\begin{equation}
\label{eq_H1}
{\bf H}({\bf k},z) = -\nabla\left[\frac{M_{\rm s}({\bf k},0)}{2k}e^{-{\bf k}z}\right],
\end{equation}
which is the magnetic field generated from the discontinuity of the magnetisation of one surface. In case of a thin film with thickness $t$ one also has to consider the magnetic field from the bottom surface. The field in the half-space above the sample then becomes
\begin{equation}
\label{eq_Htot}
{\bf H}({\bf k},z) = -\nabla\left[\frac{M_{\rm s}({\bf k},0)}{2k}e^{-{\bf k}z}(1-e^{-{\bf k}t})\right].
\end{equation}
Introducing the surface moment density $\sigma = t\cdot M_{\rm s}$ and taking the limit of a thin film ($t \rightarrow 0, 1-e^{-{\bf k}t} \rightarrow kt$), one obtains
\begin{equation}
\label{eq_H}
{\bf H}({\bf k},z) = -\nabla\left[\frac{\sigma({\bf k},0)}{2}e^{-{\bf k}z}\right].
\end{equation}

Using the above equation for the gradient, the in-plane and perpendicular components of the magnetic field read
\begin{equation}
\label{eq_Bxy}
{\bf B}_{x,y}({\bf k},z) = -i\mu_0k_{x,y}\frac{e^{-{\bf k}z}}{2}\sigma({\bf k},0) \equiv T_{x,y}\cdot\sigma({\bf k},0) 
\end{equation}
\begin{equation}
\label{eq_Bz}
{\bf B}_{z}({\bf k},z) = \mu_0{\bf k}\frac{e^{-{\bf k}z}}{2}\sigma({\bf k},0) \equiv T_{z}\cdot\sigma({\bf k},0) 
\end{equation}
The magnetic field along the NV-axis in Fourier space ${\bf B}_{\rm NV}$ is then given by a single propagator $T_{\rm NV}$:
\begin{equation}
\label{eq_BNV}
{\bf B}_{\rm NV}({\bf k},h_{\rm NV}) = \sin(\Theta_{\rm NV})\cos(\Phi_{\rm NV})B_x + \sin(\Theta_{\rm NV})\sin(\Phi_{\rm NV})B_y + \cos(\Theta_{\rm NV})B_z \equiv T_{\rm NV}\cdot\sigma({\bf k},0).
\end{equation}

In particular by solving Eq.(\ref{eq_BNV}) for $\sigma({\bf k},0)$ we can extract the surface magnetisation from the measured $B_{\rm NV}$ data. In this, high-frequency components, and therefore noise, are exponentially enhanced and need to be filtered out. A Hanning window\cite{Roth1989} low-pass filter is conveniently used to circumvent this problem:
\begin{equation}  
\label{eq_HanningFilter}
W(k)=
\begin{cases}
0.5\cdot[1+\cos(kh_{\rm NV}/2 )], &\text{for }  k<2\pi/h_{\rm NV}\\
0,               &\text{for }  k>2\pi/h_{\rm NV}
\end{cases}
\end{equation}
where the cut-off wavelength 2$\pi$/h$_{\rm NV}$ is here naturally set by the NV sample distance $h_{\rm NV}$.

Our reverse propagation procedure is then as follows:
\begin{enumerate}
	\item At each point (x,y) of the scan, the electron spin resonance frequency $\nu_{\rm ESR}$ is fitted and the magnetic field is extracted according to
	\begin{equation}
	\label{eq_ESR}
	B_{\rm NV} = \frac{|\nu_{\rm ESR} - D_0|}{\gamma_{\rm NV}},
	\end{equation}
	with $D_0=2870~$MHz being the zero-field splitting of the NV and $\gamma_{\rm NV}=28~$MHz/mT the electron gyromagnetic ratio.
	\item The externally applied bias magnetic field is subtracted from the data set to obtain $B_{\rm NV}(x,y,h_{\rm NV})$ generated by the sample .
	\item Optionally, the magnetic field data $B_{\rm NV}(x,y,h_{\rm NV})$ is extended outwards and decays towards zero using a Gaussian function. This is especially necessary for all data sets, which do not contain the entire magnetic flakes within the scan range.
	\item The magnetic field data $B_{\rm NV}(x,y,h_{\rm NV})$ is 2D Fourier transformed to obtain $B_{\rm NV}({\bf k},z)$. Zero-padding is used to the extent that all resolvable frequencies are well sampled.
	\item The magnetic field is transformed to obtain the surface magnetisation using
	\begin{equation}
	\label{eq_surfaceMag}
	\sigma({\bf k},0) = \frac{W({\bf k})\cdot B_{\rm NV}({\bf k},h_{\rm NV})}{T_{\rm NV}(h_{\rm NV},\Theta_{\rm NV},\Phi_{\rm NV})}
	\end{equation}
	with an initial guess for $h_{\rm NV}$, $\Theta_{\rm NV}$ and $\Phi_{\rm NV}$.
	\item The data is transformed to real space and cropped to its original size.
	\item The domain boundaries of $\sigma(x,y,0)$ are found and a homogeneous magnetisation $\sigma_0$ is assumed with $\sigma_{\rm approx}(x,y,0) = \sigma_0\cdot \textrm{sign}(\sigma(x,y,0))$. Then, $\sigma_{\rm approx}(x,y,0)$ is forward-propagated using the formalism described above, and compared with the original magnetic field data. Using a least square fitting routine, values for $h_{\rm NV}$, $\Phi_{\rm NV}$ and $\sigma_0$ are found which reproduce the measured magnetic field best.
	\item The values for $h_{\rm NV}$, $\Phi_{\rm NV}$ and $\sigma_0$ determined in step 7. are used to obtain the final moment density profiles $\sigma(x,y,0))$ using the reverse propagation described by Eq.(\ref{eq_surfaceMag}).
	
	\newpage
	\section{Non-invasiveness of imaging method}
	\label{noninvasiveness}
	
	\begin{figure}[h!]
		\includegraphics[width = 12cm]{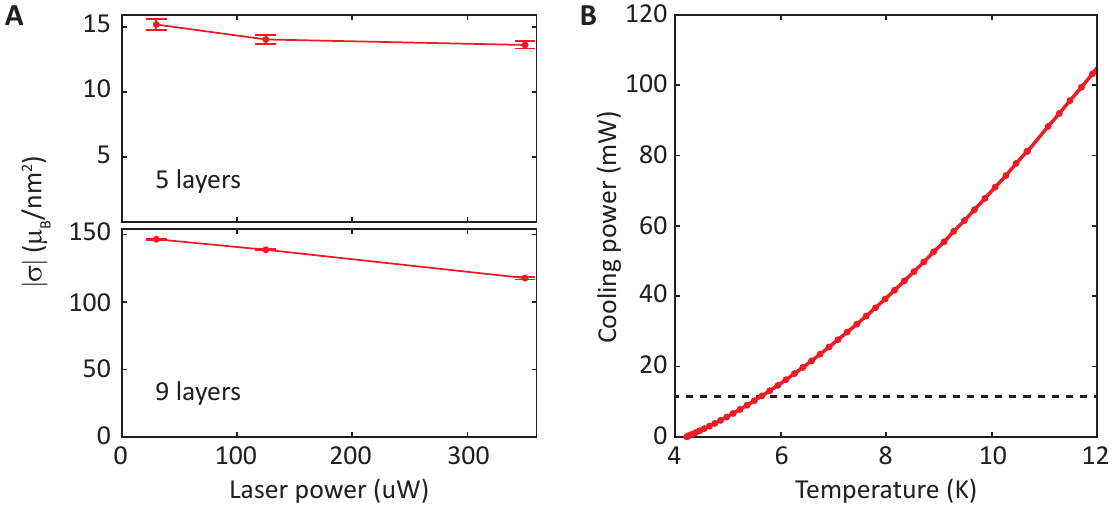}
		\caption{{\bf Impact of laser irradiation and microwave heating.} Fig.\,{\bf A} shows the influence of $532~$nm laser power on the magnetisation extracted from linecuts of B$_{\rm NV}$ across the edges of a flake. The top (bottom) panel shows magnetisation of linecuts as indicated in {\bf Fig.\,2B} ({\bf Fig.\,2C}). The linecuts over the 9-layers flake edge were performed after the puncture. Laser powers above $100~\mu$W slightly lower the value of measured magnetisation compared to the results obtained at lower powers. Working with $P_{\rm laser} \approx 40~\mu$W in all data sets ensures that no significant back-action from heating by laser power was induced onto the samples. In {\bf B} we compare the cooling power of our $^4$He bath cryostat with the power input of the microwave field required to drive electron spin resonance on the NV center. The estimated microwave power arriving in the cryostat is indicated by the black dashed line. The temperature dependent cooling power of the cryostat is shown with the red dotted line. Heating due to microwave irradiation leads to a measurement temperature of $7~$K, which is well below the Curie temperature of monolayer CrI$_3$ ($T_c=45~$K\cite{Huang2017a}).}
		\label{Fig_Invasiveness}
	\end{figure}
	
\end{enumerate}

%
%
%
%
%
%
%

\end{document}